\documentclass[a4paper,12pt]{article}
\usepackage{amsfonts,amsmath,epsfig}

\begin{document}
\title{\noindent \textbf{Rayleigh waves in anisotropic crystals
rotating about the normal to a symmetry plane}}

\author{Michel Destrade}
\date{2004}
\maketitle  

\bigskip

%
\begin{abstract}

The propagation of surface acoustic waves in a rotating 
anisotropic crystal is studied.
The crystal is monoclinic and cut along a plane containing the normal
to the symmetry plane; this normal is also the axis of rotation.
The secular equation is obtained explicitly using the ``method of the 
polarization vector'', and it shows that the 
wave is dispersive and decelerates with increasing rotation rate.
The case of orthorhombic symmetry is also treated.
The surface wave speed is computed for 12 monoclinic and 8 rhombic 
crystals, and for a large range of the rotation rate/wave frequency
ratio.

\end{abstract}
 
 \newpage
 

\section{Introduction}

Introduced more than thirty years ago, Surface Acoustic Wave (SAW) 
devices have been used with great success by the telecommunication
industry: nowadays, they are produced in large quantities (several 
billions per year) and used in wireless transmission and reception
technology for color television sets, cell phones, global positioning
systems, etc.
In recent years, new applications for SAW devices have emerged, 
namely acoustic sensors, which are passive (no power supply is needed),
resistant, almost non-aging, cheap (only one photo-lithographic process
is involved in the production), light (less than 1g) and can be 
operated remotely and wirelessly.
For instance \cite{RSOH98} SAW identification tags are used for highway
toll collection in Norway and for the Munich subway system;
SAW temperature sensors can achieve a resolution of 0.02$^\text{o}$C
from $-196^\text{o}$C up to $500^\text{o}$C;
wide ranges and fine resolutions are also achieved for pressure,
torque, or current sensors; etc.
Also, the automotive industry is engaged in the search for an 
``intelligent tire'' which could provide direct information on its 
current state as the car is moving;
in this context SAW sensors have been used to measure tire pressure 
\cite{PORS97} or friction \cite{PoSR99} as the wheel rotates.
In general, SAW devices may be used as angular rate sensors 
(gyroscopes) to measure frequency shifts due to the rotation
\cite{ClBu94a, ClBu94b, FaYJ00}.
In the present paper, an investigation of the effect of rotation upon
the speed of surface (Rayleigh) waves in an anisotropic crystal is
presented.

The crystal may possess as little as a single plane of symmetry. 
It is cut along any plane containing the normal to the symmetry plane 
and is assumed to rotate at a constant rate about this normal.
The surface wave is polarized in the symmetry plane.
In other words, it suffices to consider the propagation of a surface
wave in the $x_1$ direction of a monoclinic crystal with symmetry
plane at $x_3=0$, cut along the $x_2=0$ plane, and rotating about the
$x_3$-axis (see Fig. 1).
The secular equation for rotating materials was obtained by others 
but in simpler settings:
by Clarke and Burdess in an isotropic material, first for small 
rotation rate/wave frequency ratios \cite{ClBu94a}, then for any 
ratio \cite{ClBu94b};
by Grigor'evski\u{i}, Gulyaev, and Kozlov \cite{GrGK00} also for 
isotropic materials but neglecting the centrifugal force;
and by Fang, Yang, and Jiang \cite{FaYJ00} for crystals having 
tetragonal symmetry.
Here, the analysis is fully developed for crystals with a single 
symmetry plane, up to the derivation of the secular equation in 
explicit form, that is an equation giving the Rayleigh wave speed in 
terms of the elastic parameters and of the rotation rate.

The equation is reached in Section 3, after the governing equations
have been written down in Section 2.
The secular equation turns out to be a polynomial of degree 8 for the
squared wave speed and also for the squared rotation rate/wave 
frequency ratio. 
In the simpler case of orthorhombic symmetry (Section 4), the 
polynomial is of degree 6. 
The Rayleigh wave speed is computed numerically for 20 specific 
anisotropic materials (12 monoclinic, 8 orthorhombic) and for a 
rotation rate/wave frequency ratio varying from 0 to 10.
Of course, this range is way beyond the elastic behavior limit, and is 
irrealistic for pratical purposes where 
the frequency of a SAW device is typically in the 100kHz-10MHz range.
It is presented to show that the method of resolution is exact and not 
approximate, applies for any rate of rotation, and that in contrast 
with the non-rotating case, the secular equation is dispersive. 
At small rotation rates, and for certain crystals such as PZT-5, 
other papers \cite{FaYJ00, Coll03} show that the Rayleigh wave speed 
may at first increase slightly with the rotation frequency/wave 
frequency ratio. 
At large ratios, it is seen here that the wave speed decreases 
with increasing ratios.
These variations are crucial to the understanding and correct 
design of rotating SAW sensors or SAW signal processing devices.
A recent article \cite{JSXV02} describes the manufacturing of a 
1 cm $\times$ 1 cm SAW gyroscope and how the rotation rate may be 
measured using SAW technology.
Another example that springs to mind is that of ``spinning missiles'' 
\cite{JaHo93} for which it is reasonnable to speculate that the 
communication is ensured via SAW generation and processing of 
high-frequency signals modified by the rotation. 
Finally in Section 5, the merits of several methods of derivation for
the secular equation in non-rotating crystals are discussed.
This paper aims to provide a theoretical and analytical framework 
for the study of surface acoustic waves in rotating crystals.

\section{Basic Equations}

We consider a half-space $x_2 \ge 0$ occupied by a homogeneous 
anisotropic crystal possessing one plane of symmetry at $x_3=0$, and
rotating at a constant angular velocity $\Omega$ about the $x_3$-axis. 
We study the propagation of a surface (Rayleigh) wave in the 
$x_1$-direction, with attenuation in the $x_2$-direction.
In the rotating Cartesian frame 
$(Ox_1,Ox_2,Ox_3) \equiv (O, \mathbf{i}, \mathbf{j}, \mathbf{k})$,
the equations of motion are \cite{ScCe73}
\begin{equation} \label{motion}
\text{\textbf{div }} \mbox{\boldmath $\sigma$} = 
\rho \mathbf{u}_{,tt}
 + 2 \rho \Omega \mathbf{k} \times \mathbf{u}_{,t}
 + \rho \Omega^2 \mathbf{k} \times(\mathbf{k} \times \mathbf{u}),
\end{equation}
where $\mbox{\boldmath $\sigma$}$ is the Cauchy stress tensor,
$\rho$ is the constant mass density of the material, 
and the comma denotes differentiation.
The second term in the right hand-side of Eq.\eqref{motion} 
is due to  the Coriolis acceleration, the third is due to  the 
centrifugal acceleration.
Note that Eq.\eqref{motion} represents the time-\textit{dependent} 
part of the full equations of motion.
The time-\textit{independent} part, namely 
\textbf{div} $\mbox{\boldmath $\sigma$}^s = 
 \rho \Omega^2 \mathbf{k} \times
             [\mathbf{k} \times (\mathbf{u}^s+ \mathbf{x})]$,
where $\mathbf{u}^s = \mathbf{u}^s(\mathbf{x})$ and 
$\sigma^s_{ij} = c_{ijkl}u^s_{l,k}$ must be solved 
separately.
The questions remain of (a) whether an actual time-independent 
solution exists for all $\Omega$ and if it does, 
of (b) whether the boundary conditions of a 
traction-free rotating half-space may be satisfied without 
perturbating the time-dependent boundary value problem.
These questions do not seem to have been addressed in the literature, 
but some preliminary work seem however to suggest that (a) and (b) 
may be answered positively, at least within the framework of small 
amplitude waves superimposed upon a large elastic deformation.

Now, turning back to the time-dependent equations \eqref{motion}, 
the mechanical displacement $\mathbf{u}$ is taken in the form
\begin{equation} \label{u}
\mathbf{u}(x_1, x_2, x_3, t) = \mathbf{U} (kx_2) e^{ik(x_1-vt)},
\end{equation}
showing a sinusoidal propagation with speed $v$ and wave number $k$ 
in the $x_1$-direction, and the possibility of an attenuation in the 
$x_2$-direction through the unknown function $\mathbf{U} (kx_2)$.

We wish to describe the influence of the frame rotation upon the speed
of Rayleigh waves, and to this end, we introduce the following 
quantities,
\[
X = \rho v^2, \quad \delta = \Omega/(kv) = \Omega/\omega,
\]
where $\omega$ is the real frequency of the wave.

For two-dimensional motions ($\partial\mathbf{u}/\partial x_3=0$) 
such as Eq.\eqref{u}, the anisotropy of a crystal possessing $x_3=0$ 
as a symmetry plane is described by the following strain-stress 
relationship \cite{Ting96}:
\[
\begin{bmatrix} 
   \epsilon_{11} \\
   \epsilon_{22} \\
   2\epsilon_{23} \\
   2\epsilon_{31} \\
   2\epsilon_{12} \end{bmatrix}=
\begin{bmatrix}
  s'_{11} & s'_{12} &     0   &    0     & s'_{16} \\
          & s'_{22} &     0   &    0     & s'_{26} \\
          &         & s'_{44} & s'_{45 } &    0     \\
          &         &         & s'_{55}  &    0     \\
          &         &         &          & s'_{66} 
\end{bmatrix}
\begin{bmatrix} 
   \sigma_{11} \\
   \sigma_{22} \\
   \sigma_{23} \\
   \sigma_{31} \\
   \sigma_{12} \end{bmatrix},
\]
where  the strain components $\epsilon_{ij}$ are defined in terms of 
the displacement components by: $2 \epsilon_{ij}= u_{i,j}+u_{j,i}$, 
and the $s'_{ij}$ are the reduced compliances.
Alternatively, the equivalent strain-stress relations can be used
\cite[p.39]{Ting96},
\begin{equation} \label{StressStrain}
\mbox{\boldmath $\sigma^o$} 
  = \mathbf{C^o}\mbox{\boldmath $\epsilon^o$},
\quad
\mathbf{C^o}\mathbf{s'} = \mathbf{1},
\end{equation}
where $\mbox{\boldmath $\sigma^o$} = 
[\sigma_{11},\sigma_{22},\sigma_{23},\sigma_{31},\sigma_{12}]^\text{T}$
, $\mbox{\boldmath $\epsilon^o$} = 
[\epsilon_{11}, \ldots, 2\epsilon_{12}]^\text{T}$. 
The $C^o_{ij}$ are elements, in the Voigt notation, of the 
fourth-order elastic stiffness tensor $C_{ijkl}$.
Table 1 shows the relevant reduced compliances of 12 different 
monoclinic crystals, computed from the corresponding stiffnesses 
as collected by Chadwick and Wilson \cite{ChWi92};
the last column gives the corresponding Rayleigh wave speed in the 
non-rotating case \cite{Dest01}.

In view of the form Eq.\eqref{u} for the displacement, we introduce 
the functions $t_1$, $t_2$ for the tractions $\sigma_{12}$, 
$\sigma_{22}$ on the planes $x_2$=const. as,
\[
\sigma_{12}(x_1,x_2,x_3,t)= ik t_1(kx_2)e^{ik(x_1 -vt)},
\quad \sigma_{22}(x_1,x_2,x_3,t)= ik t_2(kx_2)e^{ik(x_1 -vt)}.
\]
Then, substituting Eqs.\eqref{u} and \eqref{StressStrain} into the
equations of motion \eqref{motion}, we derive the following system of
linear first-order differential equations for $U_1$, $U_2$, $t_1$,
$t_2$,
\begin{equation} \label{Stroh}
\begin{bmatrix}
 \mathbf{U}' \\
 \mathbf{t}' \end{bmatrix}
= i
\begin{bmatrix}
 \mathbf{N_1} & \mathbf{N_2} \\
 \mathbf{\check{N}_3} + (1 + \delta^2)X \mathbf{1}
        &  \mathbf{N_1}^{\mathrm{T}}
\end{bmatrix}
\begin{bmatrix}
 \mathbf{U} \\
 \mathbf{t} \end{bmatrix},
\end{equation}
where $\mathbf{U}= [ U_1, U_2]^{\mathrm{T}}$, 
$\mathbf{t}= [ t_1, t_2  ]^{\mathrm{T}}$,
and the prime denotes differentiation with respect to $k x_2$.
Note that, as in the static case \cite{Stro62}, 
the anti-plane strain (stress) decouples from the plane 
strain (stress) and need not be considered for this problem.
This decoupling would not occur if the the crystal was rotating about
the $x_1$-axis or the $x_2$-axis \cite{FaYJ00}.

The surface $x_2=0$ is free of tractions and so, the boundary 
conditions are
\begin{equation} \label{BC}
t_1(0) = t_2(0) = 0.
\end{equation}

In Eq.\eqref{Stroh}, $\mathbf{N_1}$ and $\mathbf{N_2}$ are the same
as the $2 \times 2$ submatrices of the $6 \times 6$ fundamental 
elasticity matrix $\mathbf{N}$ from Ingebrigsten and Tonning 
\cite{InTo69}.
Their real matrix $\mathbf{N_3}$ however has been modified by the
introduction of off-diagonal pure imaginary terms.
Explicitly, we have
\[
\begin{array}{c}
-\mathbf{N_1} = 
   \begin{bmatrix}
     r_6 & 1 \\
     r_2 & 0
    \end{bmatrix},
 \quad
\mathbf{N_2} = 
\begin{bmatrix}
     n_{66} & n_{26} \\
     n_{26} & n_{22} 
    \end{bmatrix}, 
 \quad
-\mathbf{\check{N}_3} = 
\begin{bmatrix}
    \eta     &  2i\delta X \\
 -2i\delta X &  0
    \end{bmatrix}, 
\end{array}
\]
where the quantities $r_2$, $r_6$, $n_{22}$, $n_{26}$, $n_{66}$, 
$\eta$ are given in terms of the elastic parameters as 
\cite{Dest01, Ting02a}
\begin{align} \label{coefficients}
&\eta =  \frac{1}{s'_{11}},
\quad
r_6 =  -\frac{s'_{16}}{s'_{11}},
\quad
r_2 =  -\frac{s'_{12}}{s'_{11}},
\nonumber \\
& n_{66} =\frac{1}{s'_{11}}\begin{vmatrix}
     s'_{11} & s'_{16} \\
     s'_{16} & s'_{66}
    \end{vmatrix},
\quad
n_{22} =  \frac{1}{s'_{11}}\begin{vmatrix}
     s'_{11} & s'_{12} \\
     s'_{12} & s'_{22}
    \end{vmatrix},
\nonumber  \quad
    n_{26} =  \frac{1}{s'_{11}}\begin{vmatrix}
     s'_{11} & s'_{16} \\
     s'_{12} & s'_{26}
    \end{vmatrix}.  \nonumber
\end{align}
Thus the rotation of the crystal perturbs the equations of motion in 
three ways: 
the introduction of dispersion through $\delta$;
a shift of magnitude $\delta^2$  in $X=\rho v^2$ for the lower left 
submatrix of $\mathbf{N}$ proportional to the $2 \times 2$ unit matrix;
and the modification of $\mathbf{N_3}$, which is diagonal in the 
non-rotating case
(note that the new matrix $\mathbf{\check{N}_3}$ is Hermitian:
$\overline{\mathbf{\check{N}_3}} = \mathbf{\check{N}_3}^\text{T}$.)
Despite these modifications, the secular equation can be obtained 
explicitly for the surface wave speed, using a method proposed by 
Currie \cite{Curr79} and by Taziev \cite{Tazi89} for non-rotating 
anisotropic crystals with and without a plane of symmetry, 
respectively.

\section{Secular Equation}

The method of the polarization vector was first presented by Currie
\cite{Curr79} to derive the secular equation for Rayleigh waves in
the symmetry plane of monoclinic non-rotating crystals.
Then Taziev \cite{Tazi89} generalized the method to triclinic 
(no symmetry plane) crystals.
This method takes advantage of the Cayley-Hamilton theorem for the 
fundamental matrix $\mathbf{N}$, which implies that only $n$ matrices
$\mathbf{N}^k$ ($k=1,\ldots,n$) are linearly independent ($n=3$ for
monoclinic crystals, $n=5$ for triclinic crystals).
Currie used the matrices $\mathbf{N}$, $\mathbf{N}^2$, $\mathbf{N}^3$;
Taziev, the matrices $\mathbf{N}$, $\mathbf{N}^2$, $\mathbf{N}^3$, 
$\mathbf{N}^4$, $\mathbf{N}^5$. 
Recently, Ting \cite{Ting03} placed their results within the context of
the Stroh-Barnett-Lothe formalism and improved on them by showing that
the choices of $\mathbf{N}^{-1}$, $\mathbf{N}$, $\mathbf{N}^2$ for
monoclinic crystals and of $\mathbf{N}^{-2}$, $\mathbf{N}^{-1}$, 
$\mathbf{N}$, $\mathbf{N}^2$, $\mathbf{N}^3$ for triclinic crystals
lead to simpler and more explicit secular equations.
His approach is now adapted to our present context of a rotating 
crystal with one symmetry plane.
An alternative derivation, not based on the Stroh-Barnett-Lothe 
formalism, is available elsewhere \cite{Dest04, Dest04b}.

We seek solutions to the equations of motion Eq.\eqref{Stroh} 
presenting exponential decay with distance
\[ 
\mathbf{U}(kx_2) = \mathbf{a} e^{ikpx_2}, \quad
\mathbf{t}(kx_2) = \mathbf{b} e^{ikpx_2}, \quad
\Im(p)>0,
\]
where the constant vectors $\mathbf{a}$ and $\mathbf{b}$ are related 
through \cite[p.139]{Ting96}: $b_i = (C_{k1i1} + p C_{i2k2})a_k$.
Then the equations of motion Eq.\eqref{Stroh} give
\begin{equation}\label{Stroh2}
p   \begin{bmatrix}
     \mathbf{a} \\
     \mathbf{b} 
    \end{bmatrix}=
\mathbf{\check{N}} \begin{bmatrix}
     \mathbf{a} \\
     \mathbf{b} 
    \end{bmatrix},
\end{equation}
where $\mathbf{\check{N}}$ is the $4 \times 4$ matrix in 
Eq.\eqref{Stroh}.
This eigenvalue problem yields a quartic for $p$.
We limit our investigation to the subsonic range, defined as the 
greatest interval of values for $v$ where the determinant of 
$\mathbf{\check{N}}-p\mathbf{1}$ possesses two roots $p_1$, $p_2$,
with positive imaginary parts.
We call $\mathbf{a_1}$, $\mathbf{a_2}$, and 
$\mathbf{b_1}$, $\mathbf{b_2}$, the vectors $\mathbf{a}$ and 
$\mathbf{b}$ corresponding to each root.
Then the solution is of the form \cite[p.141]{Ting96}
\[
\mathbf{U} = \mathbf{A} <e^{ikp^*}> \mathbf{q},
\quad
\mathbf{t} = \mathbf{B} <e^{ikp^*}> \mathbf{q},
\quad
<e^{ikp^*}> = \text{\textbf{diag} }(e^{ikp_1x_2}, e^{ikp_2x_2}),
\]
where $\mathbf{A} = [\mathbf{a_1}, \mathbf{a_2}]$,
 $\mathbf{B} = [\mathbf{b_1}, \mathbf{b_2}]$,
and $\mathbf{q}$ is a constant vector.
Using the boundary conditions Eq.\eqref{BC}, we have at the free 
surface $x_2=0$,
\begin{equation} \label{BC2}
\mathbf{Bq} = \mathbf{0}, \quad \text{and} \quad 
\mathbf{u}(x_1,0,x_3,t)=\mathbf{a_R}e^{ik(x_1-vt)}, \quad
\mathbf{a_R}= \mathbf{Aq}.
\end{equation}
Moreover, the matrices $\mathbf{A}$ and $\mathbf{B}$ satisfy the
orthogonality condition \cite{BaLo73},
\begin{equation} \label{orthogonal}
\overline{\mathbf{B}}^\text{T}\mathbf{A}
 + \overline{\mathbf{A}}^\text{T}\mathbf{B}= \mathbf{0}.
\end{equation}

Now, the eigenrelation Eq.\eqref{Stroh2} may be generalized for any
positive or negative integer $n$ as
\begin{equation} \label{StrohN}
p^n  \begin{bmatrix}
     \mathbf{a} \\
     \mathbf{b} 
    \end{bmatrix}=
\mathbf{\check{N}}^n \begin{bmatrix}
     \mathbf{a} \\
     \mathbf{b} 
    \end{bmatrix}, \quad
\text{where} \quad
\mathbf{\check{N}}^n = \begin{bmatrix}
 \mathbf{N^{(n)}_1} & \mathbf{N^{(n)}_2} \\
 \mathbf{\check{K}^{(n)}}
        &  \mathbf{N^{(n)}_1}^{\mathrm{T}}
\end{bmatrix} \quad 
\text{(say)}.
\end{equation}
Explicitly, the elements of $\mathbf{\check{N}}^n$ are computed by 
multiplication of $\mathbf{\check{N}}$ or its inverse by itself.
For instance, $\mathbf{\check{K}^{(n)}}$ for $n=1,2,-1$ is given by
$\mathbf{\check{K}^{(1)}} 
 =  \mathbf{\check{N}_3} + X(1+\delta^2)\mathbf{1}$,
\begin{align} \label{K(2)}
& \check{K}^{(2)}_{11}
 = -2s'_{16}[1-s'_{11}(1+\delta^2)X]/s^{'2}_{11}, 
\nonumber \\
& \check{K}^{(2)}_{12}
 = [1-(s'_{11}-s'_{12})(1+\delta^2)X - 2is'_{16}\delta X]/s'_{11}, 
 = \overline{\check{K}^{(2)}_{21}}, 
\nonumber \\
& \check{K}^{(2)}_{22} = 0, 
\end{align}
and
\begin{align} \label{K(-1)}
& \check{K}^{(-1)}_{11} 
  = -[s'_{22}(1+\delta^2) 
      -(s'_{11}s'_{22}-s^{'2}_{12})(1-\delta^2)^2X]X/D, 
\nonumber \\
& \check{K}^{(-1)}_{12}
 = [s'_{26}(1+\delta^2) + 2is'_{12}\delta 
      +(s'_{12}s'_{16}-s'_{11}s'_{26})(1-\delta^2)^2X]X/D
 = \overline{\check{K}^{(-1)}_{21}}, 
\nonumber \\
& \check{K}^{(-1)}_{22} =  [1- (s'_{11}+s'_{66})(1+\delta^2)X 
      +(s'_{11}s'_{66}-s^{'2}_{16})(1-\delta^2)^2 X^2]/D, 
\end{align}
where $D$ is a real denominator common to the $K^{(-1)}_{ij}$ whose
expression is too long to reproduce and which turns out to be 
irrelevant for the derivation of the secular equation.

Now we write in turn the second vector line of Eq.\eqref{StrohN}$_1$ 
for $p_1$ and for $p_2$, and deduce
\[
\mathbf{\check{K}^{(n)}}\mathbf{A}
 +\mathbf{\check{N_1}^{(n)}}\mathbf{B}
  = \mathbf{B}\text{ \textbf{diag} } (p_1,p_2).
\]
Multiplying this equality to the left by 
$\overline{\mathbf{a_R}}^\text{T}
 = \overline{\mathbf{q}}^\text{T} \overline{\mathbf{A}}^\text{T}$
and to the right by $\mathbf{q}$, and using 
Eqs.\eqref{BC2},\eqref{orthogonal}, we conclude that 
(see \cite{Ting03} for the non-rotating case),
\begin{equation} \label{fundamental}
\overline{\mathbf{a_R}}^\text{T}\mathbf{\check{K}^{(n)}}\mathbf{a_R}
=0.
\end{equation}
At $n=-1,1,2$, and $\mathbf{a_R}=[1,\alpha]^\text{T}$ (say), three
equations follow:
\[
\begin{array}{lllllll}
& \check{K}^{(-1)}_{12}  \alpha 
 &+&\overline{\check{K}^{(-1)}_{12}} \overline{\alpha}
  &+ & \check{K}^{(-1)}_{22}\alpha \overline{\alpha}
   &=  - \check{K}^{(-1)}_{11},
\\
& \check{K}^{(1)}_{12} \alpha 
 &+& \overline{\check{K}^{(1)}_{12}} \overline{\alpha}
  &+ & \check{K}^{(1)}_{22}\alpha \overline{\alpha} 
   &=  - \check{K}^{(1)}_{11},
\\
& \check{K}^{(2)}_{12} \alpha 
 &+& \overline{\check{K}^{(2)}_{12}} \overline{\alpha}
  &&  &=  - \check{K}^{(2)}_{11}.
\end{array} 
\]
We re-arrange this system as: $F_{ik}g_k=h_i$, by introducing the 
following quantities,
\[
\begin{array}{llll}
& F_{11} = D\Re(\check{K}^{(-1)}_{12}),
 & F_{12} = D\Im(\check{K}^{(-1)}_{12}),
  & F_{13} = D \check{K}^{(-1)}_{22}, 
\\
& F_{21} = 0,  
 & F_{22} = s'_{11}\Im(\check{K}^{(1)}_{12}),
  & F_{23} =  s'_{11}\check{K}^{(1)}_{22}, 
\\
& F_{31} = s'_{11}\Re(\check{K}^{(2)}_{12}), 
 & F_{32} = s'_{11}\Im(\check{K}^{(2)}_{12}),
  & F_{33} = 0, 
\\
& g_1 = \alpha + \overline{\alpha},
 & g_2 = i(\alpha - \overline{\alpha}),
  & g_3 = \alpha  \overline{\alpha},
\\
& h_1 = -D\check{K}^{(-1)}_{11},
 & h_2 = -s'_{11}\check{K}^{(1)}_{11},
  & h_3 = -s'_{11}\check{K}^{(2)}_{11}.
\end{array} 
\]
Note that the explicit expressions for the non-dimensional quantities 
$F_{ik}$ and $h_i$ in terms of $X=\rho v^2$, $\delta = \Omega/\omega$, 
and the $s'_{ij}$ are easily read off Eqs.\eqref{K(2)},\eqref{K(-1)}.
For instance, $F_{12}=2s'_{12}\delta X$, $F_{32}=-2s'_{16}\delta X$, 
$h_2 = 1 - s'_{11}(1+\delta^2)X$, and so on.

The linear non-homogeneous system $\mathbf{Fg}=\mathbf{h}$ has a unique
solution for $\mathbf{g}$.
Introducing $\Delta = \text{det }\mathbf{F}$ and $\Delta_k$ 
($k=1,2,3$), the determinant of the matrix obtained from $\mathbf{F}$
by replacing its $k$-th column with $\mathbf{h}$, we write the solution
as $g_k = \Delta_k/\Delta$. 
But the components of $\mathbf{g}$ are related one to another through
$g_3 = (g_1/2)^2 + (g_2/2)^2$. 
This relation is the \textit{explicit secular equation for Rayleigh
waves on an anisotropic crystal rotating in its plane of symmetry},
\begin{equation} \label{secular}
\Delta_1^2 + \Delta_2^2 - 4 \Delta_3 \Delta = 0.
\end{equation}

This equation is a polynomial of degree 8 in $X=\rho v^2$, and also 
of degree 8 in $\delta^2$.
Because $\delta = \Omega / \omega$ appears only in even powers in 
the secular equation, the Rayleigh speed obtained as a root of 
Eq.\eqref{secular} does not depend on the sense of rotation.
Numerically, we find that the rotation slows the Rayleigh wave down
and that the speed is a monotone decreasing function of $\delta$.
We see this behavior on Fig. 2, where the dependence of the Rayleigh 
wave speed upon $\delta$ is shown for the 12 monoclinic crystals from 
Table 1.
The curves are arranged in the same order as in the Table,
from the slowest (diphenyl, starting at 1276 m/s) to the fastest 
(diallage, starting at 4000 m/s).

The secular equation is valid for any crystal possessing at least one 
plane of symmetry, as long as the half-space is cut along a plane
containing the normal to the plane of symmetry.
In particular, it is also valid for orthorhombic crystals when the 
plane of cut contains one of the crystallographic axes.
When this plane contains two crystallographic axes, the secular
equation factorizes and a separate treatment is required.

\section{Orthorhombic Materials}

When the material possesses three orthogonal planes of symmetry and 
the axes ($O, \mathbf{i}, \mathbf{j}, \mathbf{k}$) are aligned with 
the crystallographic axes, some compliances vanish: 
$s'_{16}=s'_{26}=0$.
Table 2 lists the values of the relevant reduced compliances for 
8 rhombic crystals, computed from the corresponding values of the 
stiffnesses as collected by Shutilov \cite{Shut88}.
The corresponding Rayleigh wave speed $v_R$ in the non-rotating case  
(last column) is found from the exact secular equation \cite{Dest03},
\[(1-Xs'_{11}) \sqrt{1-Xs'_{66}}
  - X \sqrt{s'_{11}[s'_{22} -X(s'_{11}s'_{22}-s^{'2}_{12})]}
            = 0.
\]

When the frame is rotating, the system $\mathbf{Fg}=\mathbf{h}$ 
reduces to 
\[ 
\begin{bmatrix}
   0   & F_{12} & F_{13} \\
   0   & F_{22} & F_{23} \\
F_{31} &   0    &   0    
\end{bmatrix}
\begin{bmatrix}
 g_1 \\ g_2 \\ g_3
\end{bmatrix}
=
\begin{bmatrix}
 h_1 \\ h_2 \\ 0
\end{bmatrix},
\]
where $F_{31} \ne 0$ and
\[
\begin{array}{ll}
 F_{12} = 2s'_{12}\delta X, \quad
   F_{13} = 1 - (s'_{11}+s'_{66})(1+\delta^2)X 
            + s'_{11}s'_{66}(1-\delta^2)^2 X^2, 
\\
F_{22} = -2s'_{11}\delta X, \quad
   F_{23} =  s'_{11}(1+\delta^2)X, 
\\
h_1 = [s'_{22}(1+\delta^2) 
       - (s'_{11}s'_{22}-s^{'2}_{12})(1-\delta^2)^2X]X, \quad
h_2 = 1 - s'_{11}(1+\delta^2)X.
\end{array} 
\]
From this new system of equations, we deduce that 
$g_2 = \hat{\Delta}_2/\hat{\Delta}$ and
$g_3 = \hat{\Delta}_3/\hat{\Delta}$, where
\[
\hat{\Delta} = F_{12}F_{23} - F_{22}F_{13}, \quad
\hat{\Delta}_2 = h_1F_{23} - h_2 F_{13}, \quad
\hat{\Delta}_3 = F_{12}h_2 - F_{22}h_1, 
\]
and also that $g_1 = \alpha + \overline{\alpha} = 0$,
implying that $g_2 = 2i \alpha$, $g_3= -\alpha^2= (g_2/2)^2$ as well.
This last equality is the \textit{explicit secular equation for 
Rayleigh waves on an orthorhombic crystal rotating in one plane of 
symmetry},
\[
\hat{\Delta}_2^2 - 4\hat{\Delta}_3 \hat{\Delta} = 0.
\]

This equation is a polynomial of degree 6 in $X=\rho v^2$ and in
$\delta = \Omega/\omega$.
As in the monoclinic case above, the roots are even functions of 
$\delta$. 
Numerically, the results are similar to those of the monoclinic case,
as Fig. 3 shows for the eight orthorhombic crystals of Table 2.
Again, the curves are arranged in the same order as in the Table,
from the slowest (sulfur, starting at 1628 m/s) to the fastest 
(benzophenone, starting at 4723 m/s).

\section{Concluding Remarks}

Several methods have been proposed to derive explicitly the secular 
equation for surface waves in non-rotating monoclinic crystals with
the plane of symmetry at $x_3=0$.
This Author \cite{Dest01} wrote the equations of motion as a system of
two second-order differential equations for the tractions $\mathbf{t}$:
$\widehat{\alpha}_{ik} t_k'' - i \widehat{\beta}_{ik} t_k'
 - \widehat{\gamma}_{ik} t_k = 0$, where 
\mbox{\boldmath $\widehat{\alpha}$},
\mbox{\boldmath $\widehat{\beta}$}, 
\mbox{\boldmath $\widehat{\gamma}$} are $2 \times 2$ real symmetric 
matrices. 
Then the method of first integrals \cite{Mozh95} yields the secular
equation.
The equations of motion Eq.\eqref{Stroh} may also be written in a 
similar manner for a rotating crystal, but 
\mbox{\boldmath $\widehat{\alpha}$},
\mbox{\boldmath $\widehat{\beta}$}, 
\mbox{\boldmath $\widehat{\gamma}$} become complex and the method of 
first integrals is no longer applicable as such.
Next, Ting \cite{Ting02a} assumed an exponential form for 
$\mathbf{t}(kx_2)$ and obtained the secular equation through some 
simple algebraic manipulations, taking advantage of the fact that 
$\widehat{\alpha}_{12} = \widehat{\beta}_{22} =0$;
in the rotating case however, these quantities are no longer zero.
Furs \cite{Furs97} (using the displacement field) and this Author
\cite{Dest03} (using the traction field) devised yet another method,
where the secular equation is the resultant of two polynomials;
again, having real quantities for the components of $\mathbf{N}$ is a 
crucial property, no longer true for rotating crystals.

All in all, it seems that the method of the polarization vector is 
the most appropriate for the case of a rotating crystal.
Note that a simple derivation of its main result \eqref{fundamental}, 
not relying on the Stroh formalism, was presented recently 
\cite{Dest04, Dest04b}.

\newpage

\begin{figure}
\begin{centering}
\epsfig{figure=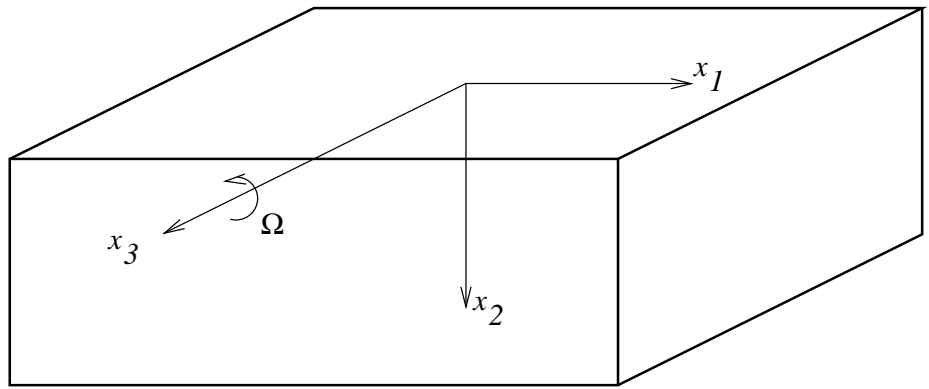}
\caption{Monoclinic crystal with symmetry plane at 
$x_3=0$, cut along $x_2=0$, and rotating about $x_3$ at constant 
angular velocity $\Omega$.}
\end{centering}
\end{figure}

\vspace*{\fill}

\newpage

\begin{center}
Table 1.
\textit{Values of the reduced compliances} ($10^{-12}$ m$^2$/N), 
\textit{density} (kg/m$^3$),
\textit{and (non-rotating) surface wave speed} (m/s) 
\textit{for 12 monoclinic crystals.}

\noindent
{\small
\begin{tabular}{l c c c c c c c c}
\hline
\rule[-3mm]{0mm}{8mm} 
material & $s'_{11}$  & $s'_{22}$     & $s'_{12}$    & $s'_{16}$
 & $s'_{26}$      & $s'_{66}$    & $\rho$     & $v_R$
\\
\hline
diphenyl        & 854  & 1858 & -366  & -698  & -1.44 & 5049
                & 1114 & 1276
\\
tin fluoride    & 345  & 228  & -59.2 & -197  & 120   & 922
                & 4875 & 1339
\\
tartaric acid   & 343  & 211  & -164  & -223  & 301   & 1650
                & 1760 & 1756
\\
oligoclase      & 133  & 227  & -108  & 97.0  & -160  & 483
                & 2638 & 2413
\\
microcline      & 94.5 & 165  & -35.1 & 47.2  & 1.69  & 446
                & 2561 & 2816
\\
gypsum          & 243  & 130  & -68.6 & 32.9  & 28.1  & 326
                & 2310 & 3011
\\
hornblende      & 63.3 & 103  & -32.7 & -15.8 & -2.72 & 320
                & 3120 & 3049
\\
aegirite-augite & 53.6 & 78.4 & -21.0 & -10.6 & -33.5 & 237 
                & 3420 & 3382
\\
epidote         & 53.3 & 49.6 & -11.3 & 17.7  & -3.74 & 237
                & 3400 & 3409
\\
augite          & 54.5 & 64.4 & -19.5 & -19.0 & -15.7 & 211 
                & 3320 & 3615
\\
diopside        & 53.1 & 58.6 & -20.1 & 24.0  & 6.98  & 186
                & 3310 & 3799
\\
diallage        & 49.8 & 69.1 & -11.3 & -6.88 & -14.5 & 166
                & 3300 & 4000
\\
 \hline
\end{tabular}
}
\end{center}

\newpage

\begin{figure}
\begin{centering}
\epsfig{figure=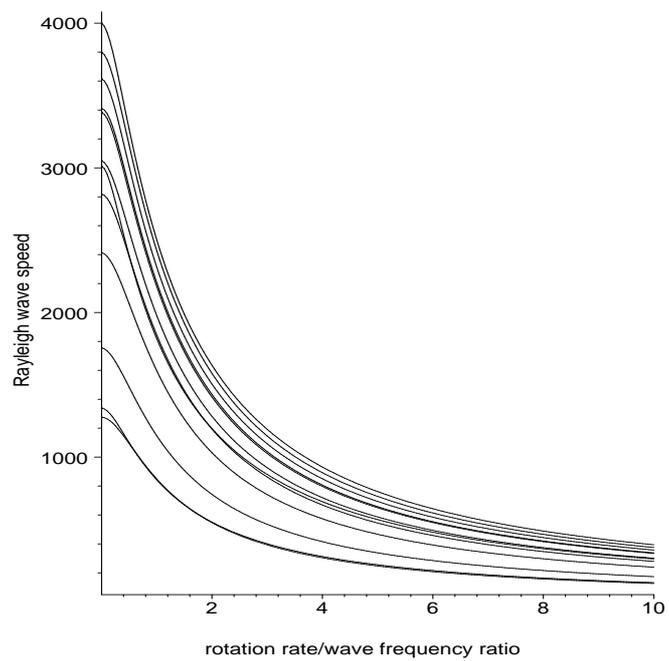, height=8.5cm, width=8.5cm}
\caption{Rayleigh wave speeds for 12 monoclinic crystals rotating 
about $x_3$.}
\end{centering}
\end{figure}

\vspace*{\fill}

\newpage

\begin{center}
Table 2.
\textit{Values of the reduced compliances} ($10^{-12}$ m$^2$/N), 
\textit{density} (kg/m$^3$),
\textit{and (non-rotating) surface wave speed} (m/s) 
\textit{for 8 orthorhombic crystals.}

\noindent
\begin{tabular}{l c c c c c c}
\hline
\rule[-3mm]{0mm}{8mm} 
material & $s'_{11}$ & $s'_{22}$ & $s'_{12}$ & $s'_{66}$ 
         & $\rho$ & $v_R$
\\
\hline
sulfur              & 65.1  & 76.2 & -42.2 & 132
                    & 2070  & 1628
\\
iodic acid          & 36.1  & 20.1 & -7.88 & 57.5 
                    & 4630  & 1678
\\
$\alpha$-uranium    & 4.89  & 5.29 & -1.13 & 13.5
                    & 19000 & 1819
\\
rochelle salt       & 49.3  & 33.0 & -18.2 & 102 
                    & 1775  & 2114
\\
sodium-tartrate     & 32.1  & 27.1 & -16.8 & 102
                    & 1818  & 2197
\\
strotium formate    & 24.5  & 30.9 & -7.32 & 58.1
                    & 2250  & 2451
\\
olivine             & 3.26  & 5.34 & -0.97 & 12.6 
                    & 3324  & 4599
\\
benzophenone        & 13.0  & 13.9 & -7.17 & 27.9
                    & 1219  & 4723
\\
 \hline
\end{tabular}
\end{center}

\newpage

\begin{figure}
\begin{centering}
\epsfig{figure=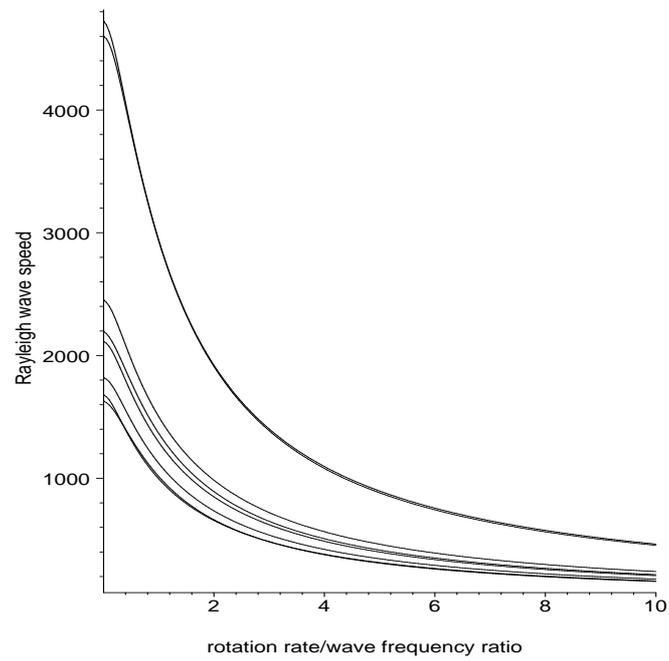,height=8.5cm,width=8.5cm}
\caption{Rayleigh wave speeds for 8 rhombic crystals rotating 
about $x_3$.}
\end{centering}
\end{figure}

\vspace*{\fill}


\end{document}